\RequirePackage{fix-cm}
\documentclass[twocolumn,epjc3,final]{svjour3}  
\smartqed  % flush right qed marks, e.g. at end of proof
 \usepackage{mathrsfs}
\usepackage{amsmath, txfonts}
\usepackage{graphicx,bm,booktabs,bm}
\usepackage{longtable,lscape}
\usepackage{ulem} 
\usepackage{overpic,multirow,color}
\definecolor{cover}{rgb}{0.77,0.87,0.88}
\definecolor{blueone}{rgb}{0.1,0.1,.7}
\definecolor{citec}{rgb}{0.14,0.47,0.09}
\definecolor{two}{rgb}{0.0,0.5,0.}
\definecolor{three}{rgb}{.5,.1,0.15}
\usepackage[bookmarks=true,bookmarksopen=false,plainpages=false,breaklinks=true,
   bookmarksnumbered=true,hypertexnames=false,
   filecolor=blue,urlcolor=three,menucolor=three,
   linkcolor=three,citecolor=blueone, colorlinks,
   anchorcolor=blue,runcolor=pink,frenchlinks=red
   pdfstartview=FitH,pdftitle=title,%
   pdfauthor=author]{hyperref}
              \allowdisplaybreaks[4]

\graphicspath{{Figures/}} %
\def\half{{\textstyle{1\over 2}}}

\journalname{Eur. Phys. J. C}
\begin{document}
\title{Combined analysis on nature of  $X(3960)$, $\chi_{c0}(3930)$, and $X_0(4140)$}
\author{Zuo-ming Ding \thanksref{addr1}
\and Jun He\thanksref{e1,addr1}
}                     % Do not remove
\thankstext{e1}{Corresponding author: junhe@njnu.edu.cn}
\institute{School of Physics and Technology, Nanjing Normal University, Nanjing 210094,  China\label{addr1}
}

\date{Received: date / Revised version: date}
\maketitle

\abstract{

In this work, a study of the $D_s\bar{D}_s$ interaction and its couplings to the
channels $D\bar{D}$ and $J/\psi \phi$  is performed in a quasipotential
Bethe-Salpeter equation approach. The $D_s^+{D}_s^-$ and $D^+{D}^-$ invariant
mass spectra in three-body $B$ decays are investigated in order to understand
the origin of  $X(3960)$, $\chi_{c0}(3930)$, and $X_0(4140)$ structures
reported at LHCb. With the help of effective Lagrangians, the
potential kernel can be constructed with meson exchanges, from which the
scattering amplitudes can be obtained. By inserting it into the three-body decay
processes, the invariant mass spectra can be calculated with an additional
Breit-Wigner resonance introduced.  The $D_s^+{D}_s^-$ invariant mass spectrum
in the  decay process $B^+\to D^+_sD^-_sK^+$ is well reproduced, and the
$X(3960)$ structure can be explained as a molecular state from the
$D_s\bar{D}_s$ interaction. The state also exhibits as a peak in the $D^+{D}^-$
invariant mass spectrum in the decay $B^+\to D^+D^-K^+$, however, it is too
narrow to reproduce the experimental structure around 3930~MeV. The dip around 4140~MeV  in the
$D_s^+{D}_s^-$ invariant mass spectrum can be reproduced by the additional
resonance with interference effect. However, a small bump instead of a dip is
produced from the $D_s\bar{D}_s-J/\psi \phi$ coupled-channel effect, which
suggests that the $X_0(4140)$ cannot be interpreted as the $D_s\bar{D}_s-J/\psi
\phi$ coupling in the current model. 

} %end of abstract

\section{Introduction}\label{sec1}

Recently, the LHCb collaboration reported two near-threshold structures named
$X(3960)$ and $X_0(4140)$ in the $D_s^+D_s^-$ invariant mass spectrum of
the decay $B^+\to D_s^+D_s^-K^+$. The mass, width and  quantum numbers of
these two structures were measured to be: $M_{X(3960)} = 3956 \pm5 \pm10$~MeV,
$\Gamma_{X(3960)} = 43  \pm13 \pm8$~MeV, $M_{X_0(4140)} = 4133 \pm 6 \pm
6$~MeV, $\Gamma_{X_0(4140)} = 67 \pm 17 \pm 7$~MeV, and both with spin parity
and charge parity $J^{P C} = 0^{++}$~\cite{LHCb:2022vsv}. The resonance peak of
$X(3960)$ is very close to the $D_s^+D_s^-$  threshold, making it a good
candidate of a $D_s^+D_s^-$ hadronic molecule. As suggested by the LHCb
collaboration, the $X_0(4140)$ might be caused by either a new resonance or by the
$D_s^+D_s^--{J/\psi\phi}$ coupled-channel effect, but a determinative
conclusion have not been drawn~\cite{LHCb:2022vsv}.

The $X(3960)$ is certainly not the first state observed in the energy region of
3.9-4.2 GeV. Over the past few decades, several experimental candidates of
charmonium-like states have been observed, such as the $X(3915)$, the
$\chi_{c0}(3930)$ and the $\chi_{c2}(3930)$. The $X(3915)$ was originally 
observed at Belle in its $\omega J/\psi$ decay mode, and was produced in both
$B$ decay~\cite{Belle:2004lle} and $\gamma\gamma$
collisions~\cite{Belle:2009and} with $J^{P C}$ determined as
$0^{++}$~\cite{BaBar:2012nxg}. In 2020, the LHCb collaboration reported the
${D}^+{D}^-$ decay mode of the $\chi_{c0}(3930)$ and $\chi_{c2}(3930)$ using $B$
decays and determined their $J^{P C}$ to be $0^{++}$ and $2^{++}$,
respectively~\cite{LHCb:2020pxc}. In the current version of the Review of
Particle Physics (PDG), the $X(3915)$ decaying to $\omega J/\psi$ and the
$\chi_{c0}(3930)$ decaying to ${D}^+{D}^-$ are assumed to be both the
$\chi_{c0}(3915)$ with $0^{++}$~\cite{ParticleDataGroup:2022pth}. The $X(3960)$
seems to be the same particle as $\chi_{c0}(3930)$ if one considers that  the
mass and width of the $X(3960)$ state are consistent with those of the
$\chi_{c0}(3930)$ meson within 3$\sigma$ measured by the LHCb
Collaboration~\cite{LHCb:2022vsv,LHCb:2020pxc}.  However, such assumption lead
to a partial width ratio ${\Gamma(X \to {D}^+{D}^-)}/{\Gamma(X \to
D_s^+D_s^-)}=0.29\pm0.09\pm0.10\pm0.08$. It contradicts the expectation that
$\Gamma(X \to {D}^+{D}^-)$ should be considerably larger than $\Gamma(X \to
D_s^+D_s^-)$ if $X$ does not have any intrinsic $s\bar{s}$
content~\cite{LHCb:2022vsv}. Hence, the $X(3960)$ and $\chi_{c0}(3930)$ are
neither the same resonance, nor the same non-conventional charmonium-like state.
It is also puzzled to denote $X(3960)$ as a conventional charmonium-like state
because the mass of $\chi_{c0}(3P)$ is around 4131-4292~MeV, which is far away
from $X(3960)$. 

Many theoretical efforts have been made into understand the origin and structure
of the $X(3960)$. Since its mass is close to the $D_s^+D_s^-$ threshold, the
hadronic molecular state is a promising picture to explain the $X(3960)$. In
Refs.~\cite{Mutuk:2022ckn,Xin:2022bzt}, the molecular state interpretation was
proposed to assign the $X(3960)$ as a $D_s^+D_s^-$ state with $J^{PC} = 0^{++}$
in the QCD sum rules approach.  In Ref.~\cite{Ji:2022uie},  the $X(3960)$ can be
well described by either a bound or a virtual state below the $D_s^+D_s^-$
threshold. Similar conclusion can be found in Ref.~\cite{Xie:2022lyw}, where
calculation was performed in an effective Lagrangian approach to study the
production rate of $X(3960)$ in the $B$ decays utilizing triangle diagrams and
assuming the $X(3960)$ as a bound/virtual state from the $D_s^+D_s^-$ interaction.
Coupled-channel analysis has also been proposed to reveal the nature of the
$X(3960)$ state, which are often combined with the analysis of the
$\chi_{c0}(3930)$. Bayar $et\ al.$ performed a coupled-channel calculation of
the interaction ${D}{\bar{D}}-{D_s}{\bar{D}_s}$  in the chiral unitary approach
and concluded that the $X(3960)$ in the $D_s^+D_s^-$ mass distribution and the
$X(3930)$ in the $D^+D^-$ mass distribution are the same
state~\cite{Bayar:2022dqa}.  Chen $et\ al.$ perfromed an analysis on decays
$\chi_{c0}(3930)\rightarrow {D}{\bar{D}}$, $X(3960)\rightarrow
{D_s}{\bar{D}_s}$, and $X(3915)\rightarrow J/\psi\omega$ using both a K-matrix
approach and a model of Flatte-like parameterization. It was suggested that the
$X(3960)$  has probably the mixed nature of a $c\bar{c}$ confining state and
${D_s}{\bar{D}_s}$ continuum~\cite{Chen:2023eix}.

There are also several theoretical and experimental works about the $X_0(4140)$
but its origin is still under debate. The LHCb collaboration interpreted it as a
new resonance with the $0^{++}$ assignment or the
${D_s}{\bar{D}_s}$-$J/\psi\phi$ coupled-channel effect~\cite{LHCb:2022vsv}.
Badalian $et\ al.$ considered $X_0(4140)$ and $X(3960)$ as exotic four-quark
states and  concluded that the $X_0(4140)$ is formed in the transitions
$J/\psi\phi$ into $D_s^{*+}D_s^{*-}$ and back in the framework of the extended
recoupling model~\cite{Badalian:2023qyi}. Agaev $et\ al.$ studied the process
$D_s\bar{D}_s-J/\psi \phi$ by QCD three-point sum rule analyses, and got the
conclusion that the structure $X_0(4140)$ may be interpreted as a hadronic
molecule $D_s^+D_s^-$, whereas the resonance X(3960) can be identified as a
tetraquark $[cs][\bar{cs}]$~\cite{Agaev:2023gti}.  Since different
interpretations were proposed in the literature, it is important to come up with
new ideas that can help us reveal the origin of $X_0(4140)$.

In our previous studies, the $D_s^+D_s^-$ molecular states were studied
in a quasipotential Bethe-Salpeter equation (qBSE) approach~\cite{Ding:2021igr}. Our calculation favors the existence of
hidden-heavy bound state $D_s^+D_s^-$  with $J^P=0^+$ at a value of cutoff
$\Lambda$=1.6~GeV. Inspired by the new experimental observation of $X(3960)$
and $X_0(4140)$, we will explore the $D_s\bar{D}_s$ interaction and its couplings to  channels ${D}{\bar{D}}$ and $J/\psi\phi$, and compare the theoretical results with the
experimental data to discuss the origin of the $X(3960)$,
$\chi_{c0}(3930)$ and $X_0(4140)$.

This work is organized as follows. After the introduction, the formalism of
three-body decay will be presented in Sec.~\ref{sec2}. Lagrangians used to
construct the potential kernels, the qBSE approach, and formula of the invariant
mass spectrum will be also presented. The numerical results of the corresponding
invariant mass spectrums will be given, and the origins of $X(3960)$, $\chi_{c0}(3930)$, and $X_0(4140)$ will be  discussed in Sec.~\ref{sec5}. A
summary of the whole work will be given in Sec.~\ref{sec6}.

\section{Formalism of three-body decay}\label{sec2}

\subsection{ Mechanism of three-body decay}

In this work, the three-body decay mechanism as illustrated in  Fig.~\ref{Fig:
diagram} in order to investigate the $X(3960)$/$\chi_{c0}(3930)$ resonance
structure observed in the process $B^+\to (X(3960)$ $/\chi_{c0}(3930))K^+$
$\to(D_s^+D_s^- /D^+D^-)K^+$ by LHCb collaboration. The $X(3960)$/$\chi_{c0}(3930)$ is assumed as an S-wave $D_s^+D_s^-$  molecular state. The
$B$ meson decays to $D_s^+D_s^-$ or $D^+D^-$ and $K^+$ firstly, and the
intermediate $D_s^+D_s^-$ molecule takes part in the rescattering process and
obtain final products $D_s^+D_s^-$ /$D^+D^-$ subsequently. Besides, to reveal
the nature of the structure found around 4140~MeV, the $J/\psi\phi-D_s^+D_s^-$
coupling will be also considered but only in the rescattering process (which is
not shown in the diagram) in order to be consistent with
experimental analysis~\cite{LHCb:2022vsv}.

\begin{figure}[h!]\begin{center}
\includegraphics[scale=0.65,trim=20 10 10 50,clip]{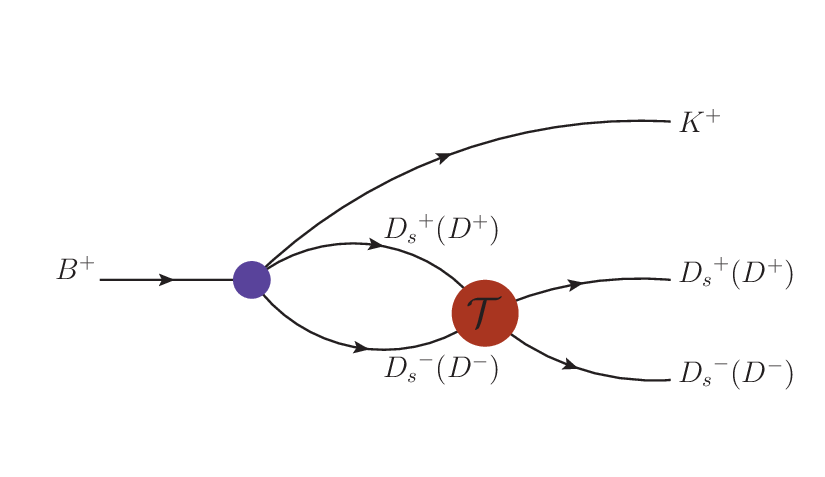}
\end{center}\caption{The diagram for the three-body decay and rescattering process of $B^+\to D_{(s)}^+ D_{(s)}^- K^+$.\label{Fig: diagram}}
\end{figure}

First, we need to treat the direct vertex $B^+\to D_{(s)}^+ D_{(s)}^- K^+$
as shown as a blue full circle in Fig.~\ref{Fig: diagram}. As in  
Ref.~\cite{Braaten:2004fk}, the amplitudes can be constrained by Lorentz
invariance to have a form of ${\cal A}_{B^+\to D_{(s)}^+ D_{(s)}^- K^+} =
c_1(c_2)$, where $c_1$ and $c_2$ are coupling constants for processes $B^+\to
{D_s}^+ {D_s}^- K^+$ and ${B^+}\to  {{D}^+} {{D}^-} {K^+}$, respectively. The
values of $c_1$ and $c_2$ can be estimated by the decay width of each process.
It can be obtained  by multiplying their branching fractions $
\mathcal{B}_{{B^+}\to  {{D}^+} {{D}^-} {K^+}} = (2.2 \pm 0.7 ) \times
{10^{-4}}$ and $ \mathcal{B}_{B^+\to D_s^+ D_s^- K^+} = (1.15 \pm 0.07 \pm 0.06 \pm
0.38) \times {10^{-4}}$~\cite{LHCb:2022dvn} with the decay width of $B$ meson
obtained from its mean life $\Gamma_B = {1}/{\tau} = 6.105 \times {10^{11}}
~s^{-1} = 4.017  \times {10^{-10}} $~MeV~\cite{ParticleDataGroup:2022pth}. The
values of $c_1$ and $c_2$ can be obtained as $c_1 = 6.027 \times 10^{-5}$ and
$c_2 = 5.397 \times 10^{-5}$. Here the rescattering effect is not considered, which effect will be discussed later.

\subsection{Flavor wave functions and Lagrangians involved}

The  potential kernel of rescattering process as
shown in Fig.~\ref{Fig: diagram} should be constructed with effective Lagrangians
and flavor wave functions to find the pole in the complex energy plane within qBSE approach and to calculate
the invariant mass spectrum. In Ref.~\cite{He:2014nya}, it was
explained explicitly how to construct a potential for states with definite
isospins under SU(3) symmetry with the corresponding flavor wave functions. For
the S-wave ${D_s}^+{D_s}^-$ and ${D^+}{D^-}$ states, the wave functions can be
constructed as \begin{align}
|X_{D\bar{D}}^0\rangle_{I=0}&=\frac{1}{\sqrt{2}}\left(|D^{+}\bar{D}^{-}\rangle+|D^0\bar{D}^{0}\rangle\right),
|X_{D_{s}\bar{D}_{s}}^0\rangle=|D_{s}^{-}D_{s}^{+}\rangle.  \label{wf} \end{align}

The meson exchanges were adopted to achieve the potential kernel for the
interactions considered. According to the chiral  and heavy quark
symmetries, the Lagrangians for heavy mesons interacting with light mesons
read\cite{Cheng:1992xi,Yan:1992gz,Wise:1992hn,Burdman:1992gh,Casalbuoni:1996pg}
\begin{align} \mathcal{L}_{\mathcal{PP}\mathbb{V}} &=
-\sqrt{2}\beta{}g_V\mathcal{P}^{}_b\mathcal{P}_a^{\dag} v\cdot\mathbb{V}_{ba}
+\sqrt{2}\beta{}g_V\widetilde{\mathcal{P}}^{\dag}_a \widetilde{\mathcal{P}}^{}_b
v\cdot\mathbb{V}_{ab},\label{Eq:V}\\
\mathcal{L}_{\mathcal{PP}\sigma} &=
-2g_s\mathcal{P}^{}_b\mathcal{P}^{\dag}_b\sigma
-2g_s\widetilde{\mathcal{P}}^{}_b\widetilde{\mathcal{P}}^{\dag}_b\sigma,\label{Eq:sigma}
\end{align} where ${\mathcal{P}}^{T} =(D^{0},D^{+},D_s^{+})$. The velocity $v$
should be replaced by $i\overleftrightarrow{\partial}/2\sqrt{m_im_f}$ with the
$m_{i,f}$ being the mass of the initial or final heavy meson. The $\mathbb V$
denotes the vector matrix as \begin{align} \mathbb{V}=\left(\begin{array}{ccc}
\frac{\rho^0+\omega}{\sqrt{2}}&\rho^{+}&K^{*+}\\
\rho^{-}&\frac{-\rho^{0}+\omega}{\sqrt{2}}&K^{*0}\\ K^{*-}&\bar{K}^{*0}&\phi
\end{array}\right).\label{MPV} \end{align} The parameters involved here were
determined in the literature as $\beta=0.9$, $g_s=0.76$ and $g_V = 5.9$
~\cite{Falk:1992cx,Isola:2003fh,Liu:2009qhy,Chen:2019asm}. 

In Refs.~\cite{Aceti:2014uea,He:2015mja}, it was suggested that the $J/\psi$ exchange is important in the $D\bar{D}^*$ interaction to reproduce the $Z_c(3900)$.  In this work, the couplings of heavy-light charmed mesons to $J/\psi$ are also considered and written with the help of heavy quark effective theory as \cite{Casalbuoni:1996pg,Oh:2000qr},
\begin{align}
{\cal L}_{D_{(s)} \bar{D}_{(s)}J/\psi} &=
ig_{D_{(s)}D_{(s)}\psi} \psi \cdot
\bar{D}\overleftrightarrow{\partial}D,\label{psi}
\end{align}
where the couplings are related to a single parameter $g_2$ as ${g_{D_{(s)}D_{(s)}\psi}}/{m_D}= 2 g_2 \sqrt{m_\psi }$, with $g_2={\sqrt{m_\psi}}/({2m_{D_{(s)}}f_\psi})$ and $f_\psi=405$~MeV. 

To investigate whether the structure found around 4140 MeV is caused by the
$J/\psi\phi - {D_s}^+{D_s}^-$ coupled-channel effect, which should be taken into
consideration in rescattering process.  Pseudoscalar $\bar{D}_s$ and ${D}_s$
mesons and vector $\bar{D}^*_s$ and ${D}^*_s$ mesons are exchanged during the
coupling process $J/\psi\phi - {D_s}^+{D_s}^-$. To describe this
process, apart from Eq.~(\ref{Eq:V}) and Eq.~(\ref{psi}), we also need the
Lagrangians as~\cite{Casalbuoni:1996pg,Oh:2000qr}
\begin{align}
{\cal L}_{D^*_{(s)}\bar{D}_{(s)}J/\psi}&=
-g_{D_{(s)}^*D_{(s)}\psi} \,  \, \epsilon_{\beta \mu \alpha \tau}
\partial^\beta \psi^\mu (\bar{D}
\overleftrightarrow{\partial}^\tau D^{* \alpha}+\bar{D}^{* \alpha}
\overleftrightarrow{\partial}^\tau D),\nonumber\\
\end{align}
where ${g_{D_{(s)}^*D_{(s)}^*\psi}}/{m_{D^*}} = 2 g_2 \sqrt{m_\psi }$, and the value of $g_2$ have been mentioned under Eq.~(\ref{psi}). 

\subsection{Potential kernel}

The potential  interaction can be constructed by the meson exchange as~\cite{Ding:2021igr,He:2019ify,Ding:2020dio},
\begin{align}%
{\cal V}_{\mathbb{P},\sigma}=I_i\Gamma_1\Gamma_2 P_{\mathbb{P},\sigma}f(q^2),\ \ 
{\cal V}_{\mathbb{V}}=I_i\Gamma_{1\mu}\Gamma_{2\nu}  P^{\mu\nu}_{\mathbb{V}}f(q^2),\label{V1}
\end{align}
Here the vertices $\Gamma$ can be obtained by the standard Feynman rule from the Lagrangians given in the above.  In the current work, the SU(3) symmetry is considered. Hence, with the help of the wave function in Eq.~(\ref{wf}) and the flavor part in Lagrangians and the matrix in Eq.~(\ref{MPV}),   a flavor factor can attracted for a state with a certain exchanged meson based on the Feynman diagram of the meson exchanges~\cite{Ding:2021igr,He:2014nya,He:2015mja,Ding:2020dio}. Except the flavor factor $I_{\rho}={3}/{2}$,  $I_{\omega}={1}/{2}$ for the $D\bar{D}$ state, all other flavor factor has values of 1 as $I_{\sigma}=I_{J/\psi}=I_{\phi}=I_{D_s}=I_{\bar{D}_s}=I_{{D}^*_s}=I_{\bar{D}^*_s}=1 $. The value of the flavor factor $I_{K^*}$ equals to 1 theoretically, but we take it equals to 0.3, which will be discussed later. The propagators are defined as usual as
\begin{align}%
P_{\mathbb{P},\sigma}= \frac{i}{q^2-m_{\mathbb{P},\sigma}^2},\ \
P^{\mu\nu}_\mathbb{V}=i\frac{-g^{\mu\nu}+q^\mu q^\nu/m^2_{\mathbb{V}}}{q^2-m_\mathbb{V}^2},
\end{align}
with $q$ being the momentum of the exchanged meson. A form factor $f(q^2)=\Lambda_e^2/(\Lambda_e^2-q^2)$ is introduced to compensate the off-shell effect of exchanged meson with a cutoff $\Lambda_e$.The current form factor can avoid overestimation of the contribution of $J/\psi$ exchange.

\subsection{Quasipotential Beth-Salpeter approach}
In the above, we construct the potential of the interactions considered in the current work. The rescattering amplitude can be obtained with the qBSE~\cite{He:2014nya,He:2015mja,He:2017lhy,He:2015yva,He:2017aps}. After  the partial-wave decomposition,  the qBSE can be reduced to a 1-dimensional  equation with a spin-parity $J^P$ as~\cite{He:2015mja},
\begin{align}
i\hat{\cal T}^{J^P}_{m,n}({\rm p},{\rm p}')
&=i\hat{\cal V}^{J^P}_{m,n}({\rm p},{\rm
p}')+\sum_{k}\int\frac{{\rm
p}''^2d{\rm p}''}{(2\pi)^3}\nonumber\\
&\cdot
i\hat{\cal V}^{J^P}_{m,k}({\rm p},{\rm p}'')
G_0({\rm p}'')i\hat{\cal T}^{J^P}_{k,n}({\rm p}'',{\rm
p}'),\label{Eq: BS_PWA}
\end{align}
where $\hat{\cal T}^{J^P}$ and $\hat{\cal V}^{J^P}$ are partial-wave rescattering amplitude and potential. The $n$, $m$, $k$ denotes the independent helicities $\lambda_{(2,3)}$ of two constituents for the initial, final and intermediate mesons, respectively. If two constituents have helicities $\lambda_1=\lambda_2=0$, such as the case of  the $D_s\bar{D}_s$ interaction in the current work, a factor $f=1/\sqrt{2}$ should be introduced for the partial wave amplitude $\hat{\cal T}^{J^P}$ and potential $\hat{\cal V}^{J^P}$, otherwise, $f=1$. The explicit explanations about the independent amplitudes and the factor can be find in Ref.~\cite{He:2015mja}. $G_0({\rm p}'')$ is reduced propagator with the spectator approximation in the center-of-mass frame with $P=(M,{\bm 0})$ as
\begin{align}
	G_0&=\frac{\delta^+(p''^{~2}_h-m_h^{2})}{p''^{~2}_l-m_l^{2}}\nonumber\\
	&=\frac{\delta^+(p''^{0}_h-E_h({\bm p}''))}{2E_h({\bm p''})[(W-E_h({\bm
p}''))^2-E_l^{2}({\bm p}'')]}.
\end{align}
As required by the spectator approximation, the heavier particle (remarked as $h$)  is put on shell, which has  $p''^0_h=E_{h}({\rm p}'')=\sqrt{
m_{h}^{~2}+\rm p''^2}$. The $p''^0_l$ for the lighter particle (remarked as $l$) is then $W-E_{h}({\rm p}'')$ with $W$ being the total energy of the system of 2 and 3 particles. Here and hereafter we define the value of the momentum ${\rm p}=|{\bm p}|$.

The potential kernel $\hat{\cal V}_{m,n}^{J^P}/(f_mf_n)={\cal V}_{\lambda_2\lambda_3,\lambda'_2\lambda'_3}^{J^P}$ can be obtained from the potential in Eq.~(\ref{V1}) as
\begin{align}
i{\cal V}_{\lambda_2\lambda_3,\lambda'_2\lambda'_3}^{J^P}({\rm p},{\rm p}')
&=2\pi\int d\cos\theta
~[d^{J}_{\lambda_{32}\lambda'_{32}}(\theta)
i{\cal V}_{\lambda_2\lambda_3,\lambda'_2\lambda'_3}({\bm p},{\bm p}')\nonumber\\
&+\eta d^{J}_{-\lambda_{32}\lambda'_{32}}(\theta)
i{\cal V}_{\lambda_2\lambda_3,-\lambda'_2-\lambda'_3}({\bm p},{\bm p}')],
\end{align}
where $\eta=PP_2P_3(-1)^{J-J_2-J_3}$ with $P$ and $J$ being parity and spin for system and constituent 2 or 3. $ \lambda_{32}=\lambda_3-\lambda_2$. The initial and final relative momenta are chosen as ${\bm p'}=(0,0,{\rm p'})$  and ${\bm p}=({\rm p}\sin\theta,0,{\rm p}\cos\theta)$. The $d^J_{\lambda\lambda'}(\theta)$ is the Wigner d-matrix. An  exponential regularization  was also introduced as a form factor into the reduced propagator as $G_0({\rm p}'')\to G_0({\rm p}'')e^{-2(p''^2_l-m_l^2)^2/\Lambda_r^4}$ with the $m_l$ and $\Lambda_r$ being the mass of light constituent and a cutoff, respectively~\cite{He:2015mja}. The cutoff $\Lambda_r$ and the cutoff $\Lambda_e$ for exchanged meson provide similar effect on the results, hence, we take them as a unified parameter $\Lambda$.

The rescattering amplitude ${\cal T}$ can be obtained by discretizing the momenta ${\rm p}'$,
${\rm p}$, and ${\rm p}''$  in the integral equation~(\ref{Eq: BS_PWA}) by the Gauss quadrature with a weight $w({\rm
p}_i)$. After such treatment, the integral equation can be transformed to a matrix equation~\cite{He:2015mja},
\begin{align}
{T}_{ik}
&={ V}_{ik}+\sum_{j=0}^N{ V}_{ij}G_j{ T}_{jk}.\label{Eq: matrix}
\end{align}
The propagator $G$ is a diagonal matrix as
\begin{align}
	G_{j>0}&=\frac{w({\rm p}''_j){\rm p}''^2_j}{(2\pi)^3}G_0({\rm
	p}''_j), \nonumber\\
G_{j=0}&=-\frac{i{\rm p}''_o}{32\pi^2 W}+\sum_j
\left[\frac{w({\rm p}_j)}{(2\pi)^3}\frac{ {\rm p}''^2_o}
{2W{({\rm p}''^2_j-{\rm p}''^2_o)}}\right],
\end{align}
with on-shell momentum ${\rm p}''_o=\lambda^\half(W,m_2,m_3)$ with the $\lambda(x,y,z)=[x^2-(y+z)^2][x^2-(y-z)^2]$.

\subsection{Invariant mass spectrum}

With the preparation above, the invariant mass spectrum of  processes considered can be given  with direct three-body decay amplitude ${\cal A}^{J^P}_{m, \lambda_B}={\cal A}_{B^+\to {D_{(s)}}^+ {D_{(s)}}^- K^+}$ (here we remark it as $\hat{\cal A}^{J^P}_{m, \lambda_B}=f_m{\cal A}^{J^P}_{m, \lambda_B}$ with pseudoscalar $B$ meson having a helicity $\lambda_B=0$) and the rescattering amplitude ${\cal T}^{J^P}$ as Ref.~\cite{He:2017lhy},
\begin{align}
{d\Gamma\over dM_{23}}&=\frac{1}{16M}\frac{1}{(2\pi)^5}\frac{\breve{\rm p}_1{\rm p}^{cm}_3}{M}\sum_{m,\lambda_B;J^P}\frac{1}{N_J^2}|\hat{\cal M}^{J^P}_{m;\lambda_B}(M_{23})|^2,\label{Eq: IM}
\end{align}
with 
\begin{align}
\hat{\cal M}^{J^P}_{m,\lambda_B}(M_{23})
&=\hat{\cal A}^{J^P}_{m,\lambda_B}(M_{23})+\sum_{l}\int \frac{d{\rm p}'^{cm}_3{\rm p}'^{cm2}_3}{(2\pi)^3}\nonumber\\
&i\hat{\cal T}^{J^P}_{m;n}({\rm p}'^{cm}_3,M_{23}) G_0({\rm p}'^{cm}_3) \hat{\cal A}^{J^P}_{n;\lambda_B}({\rm p}'^{cm}_3,M_{23}),\label{Eq: IM0}
\end{align}
where $M$ is the mass of the $B$ meson. 
$N_J^2=(2J+1)/4\pi=1/4\pi$ with $J=0$ here. ${\rm p}^{cm}_3$ and ${\rm p}'^{cm}_3$
are the center of mass momentum in the $23$ and $2'3'$ system, which can be
expressed as $\mathrm{p}_{3^{(')}}^{c m}=
\sqrt{\lambda\left(M_{2^{(')}3^{(')}}^2, m_{3^{(')}}^{2},
m_{2^{(')}}^{2}\right)}/{2 M_{2^{(')}3^{(')}}}$. And $\breve{\rm p}_1$ satisfies
$M_{23}^2=(M-\breve{\rm E}_1)^2-\breve{\rm p}_1^2$. 

Eq.~(\ref{Eq: IM0}) can be abbreviated as matrix form $ M =  A + TG A$ by the
same discretizing  in Eq.~(\ref{Eq: matrix}), where $T$ satisfies $T =  V + VGT$.
The rescattering amplitude $T$ can be solved as  ${ T}=(1-{V}G)^{-1} V$. Since
we focus on the pole of the rescattering amplitude, we need to find the position
where $|1-{ V}G|=0$ with $z=E_R + i\Gamma/2$ corresponding  to the total energy and width at the
real axis.

\section{Numerical results and discussions}\label{sec5}

With the  above preparation, the invariant mass spectra  of $B$ decays
considered in the current work will be calculated and compared with the
experimental data. In this section, the $D_s\bar{D}_s$ interaction and its couplings to channels $D\bar{D}$ and $J/\psi\phi$ will be considered.

\subsection{$D_s^+{D}_s^-$ invariant mass spectrum with $D_s\bar{D}_s$ interaction}\label{ss1}

First, we consider the simplest model, that is, only the $D_s\bar{D}_s$
interaction is included in the rescattering  to provide a basic picture of
molecular state from the interaction. In such model, the intermediate and final
products are all $D_s\bar{D}_s$. And in the current work, we only
consider the S-wave state with spin parity $J^P$ = $0^+$. The corresponding
cutoff $\Lambda$ is adjusted to 1.8~GeV, which is a little bigger than the one
in our previous work~\cite{Ding:2021igr}, which makes the result fit better with
the new experimental data. Besides the obvious peak near the $D_s\bar{D}_s$
threshold, a structure can be seen around 4140~MeV.  Obviously, such structure can
not be produced from the $D_s\bar{D}_s$ interaction as shown in  Eq.~(\ref{Eq:
IM}). Here, as suggested by LHCb collaboration~\cite{LHCb:2022vsv}, we introduce
a Breit-Wigner resonance, which is used to fit the dip around 4140~MeV as
\begin{align}
M_{BW} &=
e^{i\theta}\cdot \frac{a\Gamma_0M_0}{M^2-{M_0}^2+i\Gamma_0M_0},\label{Eq: BW}
\end{align}
where $M_0$ and $\Gamma_0$ are the mass and width of the structure obtained
from the experimental results. The strength factor $a$ and phase angle $\theta$
are free parameters. The theoretical results for events can be obtained by
multiplying the theoretical decay distribution on efficiency, which equals to
$N_{sig}/(N_{B^+}\mathcal{B}_{B^+\to {D_s}^+ {D_s}^- K^+})$ with $N_{sig}$,
$N_{B^+}$ and  $\mathcal{B}$ being the numbers of signal events, the total
number of $B$ candidates, and the branching fraction, respectively.
Unfortunately, though the $N_{sig}$ and $\mathcal{B}$ were provided by LHCb
collaboration in Ref.~\cite{LHCb:2022vsv,LHCb:2022dvn}, the $N_{B^+}$ was not
reported, so we have to introduce a parameter multiplying on the theoretical
decay distribution to fit the experimental data. After such treatment, the
comparison between the theoretical and experimental results can be carried out. We would like to remind that such treatment make the absolute values of the coupling constants $c_1$ and $c_2$ meaningless. 

\begin{figure}[h!]\begin{center}
  \includegraphics[scale=0.9,bb=110 60 500 300,clip]{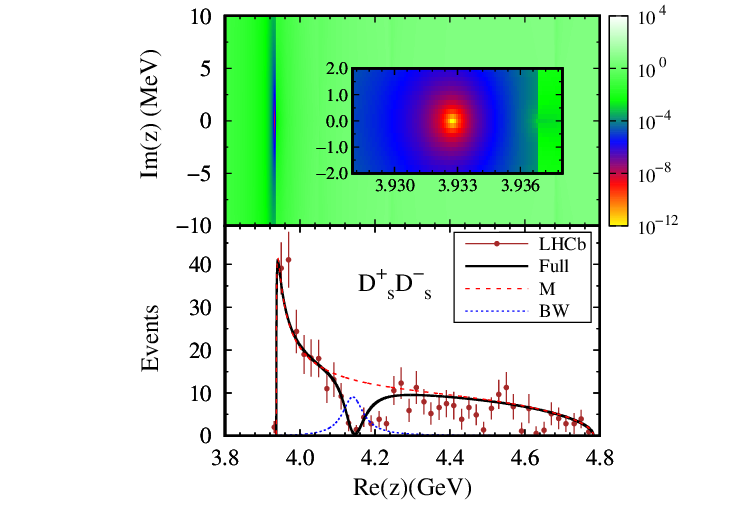}
  \end{center}\caption{The pole (upper panel) and the $D_s^+{D}_s^-$ invariant mass spectrum for decay $B^+\to D_s^+D_s^-K^+$ (lower panel) with the $D_s\bar{D}_s$ interaction. The black (full), red (dashed), and blue (dotted) curves are for the total, rescattering as ${\cal M}$ in Eq.~(\ref{Eq: IM0}), and Breit-Wigner resonance in Eq.~(\ref{Eq: BW}). The red points with error bars are the data from the LHCb experiment cited from Ref.~\cite{LHCb:2022vsv}. \label{Fig: 1}}
  \end{figure}

The pole and $D_s^+{D}_s^-$ invariant mass spectrum for decay $B^+\to
D_s^+D_s^-K^+$ are presented in Fig.~\ref{Fig: 1} and compared with the
experiment. 
In the upper panel of Fig.~\ref{Fig: 1},  a pole is found at 3932.8
GeV in the first Riemann sheet on the real axis near $D_s\bar{D}_s$ threshold,
which suggests a very weakly bound state from the $D_s\bar{D}_s$
interaction and close to the mass of $X(3960)$ measured by the LHCb
Collaboration. To confirm their relation, the invariant mass spectrum in energy
region of 3800 $-$ 4800~MeV is calculated and compared with the experiment. In
the lower panel of Fig.~\ref{Fig: 1}, a sharp peak arises near the
$D_s\bar{D}_s$ threshold, which is from the $D_s\bar{D}_s$ rescattering. With
increase of the energies, a dip can be found around 4140 MeV, which is due to
the Breit-Wigner resonance introduced in  Eq.~(\ref{Eq: BW}). We evaluate the
$\chi^2$ between the experimental data and our theoretical results in two energy regions
of 3800 $-$ 4170~MeV and of 3800 $-$ 4800~MeV. The former one is predominantly
relevant to the $X(3960)$ and $X_0(4140)$ structures, and the latter one is
relevant to all energy region.  The $\chi^2$ in energy region of
3800 $-$ 4170~MeV is equal to 13.45 for 13 data points, while the $\chi^2$ in
energy region of 3800 $-$ 4800~MeV is equal to 53.06 for 43 data points, which
indicates that our theoretical data are in a good agreement with experimental
data, and favors the assumption of the $X(3960)$ as a $D_s\bar{D}_s$ molecular
state.

\subsection{Discussion about relation between $X(3960)$ and $\chi_{c0}(3930)$}\label{ss2}

\begin{figure}[h!]\begin{center}
  \includegraphics[scale=1.2,bb=120 65 500 300,clip]{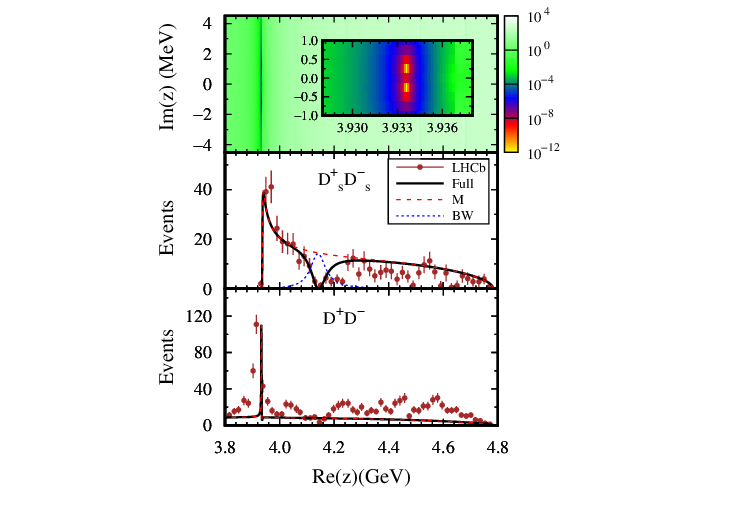}
  \end{center}\caption{ The pole (upper panel) and the$D_s^+{D}_s^-$ (middle panel) and $D^+{D}^-$ (lower panel) invariant mass spectra with the $D\bar{D}$-$D_s\bar{D}_s$ coupled-channel interaction. The black (full), red (dashed), and blue (dotted) curves are for the total, rescattering as ${\cal M}$ in Eq.~(\ref{Eq: IM0}), and Breit-Wigner resonance in Eq.~(\ref{Eq: BW}). The red points with error bars are the data of $X(3960)$ (middle panel) and $X(3930)$ (lower panel) from the LHCb experiment [24]. \label{Fig: 2}}
  \end{figure}

In this subsection, we perform a study of the $B$ meson decay with rescattering
of $D_s\bar{D}_s$-$D\bar{D}$ in order to discuss the relation between $X(3960)$
and $\chi_{c0}(3930)$. The intermediate and final products can be $D_s\bar{D}_s$
and $D\bar{D}$ with spin parity $J^P$ = $0^+$. The value of flavor factor
$I_{K^*}$ is adjusted to 0.3 instead of 1. It is interesting to found that if we
take $I_{K^*}$= 1 as its theoretical value, the $D\bar{D}$ state will remain but
the $D_s\bar{D}_s$ state will disappear, only if decreasing the value of
$I_{K^*}$ will these two states appear, which is analogous to
Ref.~\cite{Bayar:2022dqa} and may cause by SU(3) symmetry breaking for the $K^*$
meson.  A Breit-Wigner resonance is still added into the amplitude of
$D_s\bar{D}_s$-$D\bar{D}$ rescattering to match the dip found around 4140~MeV in
Refs.~\cite{LHCb:2022vsv,LHCb:2022dvn}.  The corresponding cutoff
$\Lambda$ will be also chosen as
1.8~GeV for both $D_s\bar{D}_s$
and $D\bar{D}$ channels.  The $D_s^+{D}_s^-$ and $D^+{D}^-$ invariant mass spectra are presentedn
in Fig.~\ref{Fig: 2} and compared with the experiment.

In the upper panel of Fig.~\ref{Fig: 2}, the poles leave the real axis and become
two conjugate poles in the complex energy plane after including the
coupled-channel effect. These two conjugate poles are located at
3933.6$\pm$0.25$i$~MeV, which means a molecular state with a mass of 3933.6~MeV and a width of 0.5~MeV close to the corresponding $D_s\bar{D}_s$
threshold. The $D_s^+{D}_s^-$ invariant mass spectrum  in the middle panel of
Fig.~\ref{Fig: 2} looks similar to the single-channel results in energy region of
3800 $-$ 4170~MeV in  Fig.~\ref{Fig: 1}. The $\chi^2$ between the experimental
data and our theoretical data in energy region of 3800 $-$ 4170~MeV is equal to
14.88 for 13 data points, which is close to the above single-channel results.
However, the $\chi^2$ in energy region of 3800 $-$ 4800~MeV is equal to 76.13
for 43 data points, which is a little larger than the one obtained in
subsection.~\ref{ss1}. Since in the higher energy region, the real mechanism
should be more complex, the current result is still accepted if we focus on
the structure near the $D_s\bar{D}_s$ threshold.

In the lower panel of Fig~\ref{Fig: 2}, the $D^+{D}^-$ invariant mass spectrum
are presented and compared with the data of $X(3930)$ from LHCb
experiment~\cite{LHCb:2020pxc}. As shown in Fig~\ref{Fig: 2}, the peak near
3930~MeV is extremely narrow, seems too sharp compared with the experimental
structure.  If a wider peak in the $D^+{D}^-$ invariant mass spectrum is reached,
the peak in the  $D_s^+{D}_s^-$ invariant mass spectrum will disappear. Our
calculation suggests that though peaks can be produced simultaneously near the
$D_s\bar{D}_s$ threshold in both  $D_s^+{D}_s^-$ and $D^+{D}^-$  invariant mass spectrum, the explicit shape of the experimental structure can not be well-fitted
simultaneously. In the experimental article~\cite{LHCb:2020pxc}, such structure
is suggested to be formed by two states $\chi_{c0}(3930)$ and $\chi_{c2}(3930)$,
and the $\chi_{c0}(3930)$ has a smaller width. Hence, it is still possible to
assign the $\chi_{c0}(3930)$  as a $D_s\bar{D}_s$ molecular state but with a
very small width. If so, the $\chi_{c2}(3930)$ may play a more important role to
form the structure around 3930~MeV in the $D^+{D}^-$ invariant mass spectrum.

\subsection{Discussion about origin of $X(4140)$}\label{ss3}

\begin{figure}[h!]\begin{center}
\includegraphics[scale=0.77,trim=10 30 30 40,clip]{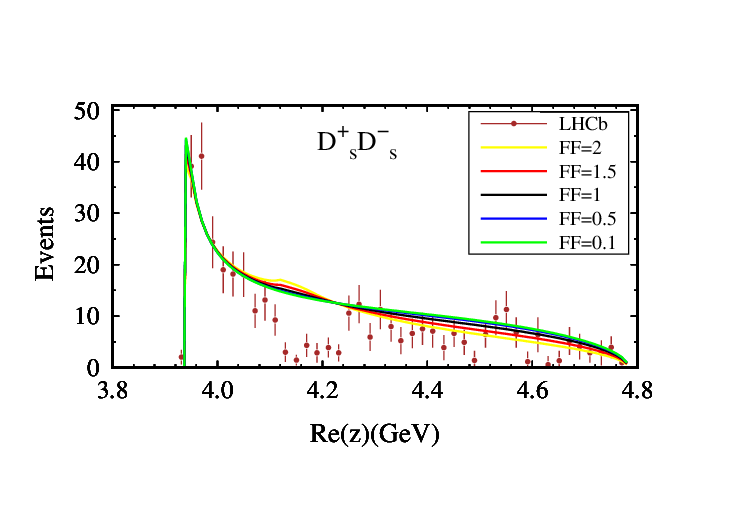}
\end{center}\caption{ The $D_s^+{D}_s^-$ invariant mass spectrum with the $D_s\bar{D}_s$-$J/\psi\phi$ coupled-channel interaction. The yellow, red, black, blue and green curves are for the mass distribution with flavor factor (FF) equal to 2, 1.5, 1, 0.5 and 0.1, respectively. The red points with error bars are the data of $X(3960)$ from the LHCb experiment ~\cite{LHCb:2022vsv}. \label{Fig: 3}}
\end{figure}

As suggested by the LHCb collaboration, the additional
structure found around 4140~MeV in the $D_s^+{D}_s^-$ invariant mass spectrum
might be caused either by a new resonance with the $0^{++}$ assignment or by a
$D_s\bar{D}_s$-$J/\psi\phi$ coupled-channel effect~\cite{LHCb:2022vsv}. In the above, we adopt the
former assumption by introducing a Breit-Wigner resonance. For better
understanding of the origin of this structure, the $D_s\bar{D}_s$-$J/\psi\phi$
coupled-channel effects with spin parity $J^P$ = $0^+$ is introduced to replace
the Breit-Wigner resonance. The corresponding cutoff $\Lambda$ will be chosen as 1.8~GeV as above calculation. We also study effect
of flavor factors  of exchanged meson $I_{D_s}$, $I_{\bar{D}_s}$, $I_{{D}^*_s}$
and $I_{\bar{D}^*_s}$ by changing their values. The result is shown in
Fig.~\ref{Fig: 3}. 

The black curve is the mass distribution obtained by theoretical values of
flavor factor of the exchanged mesons, that is,
$I_{D_s}$=$I_{\bar{D}_s}$=$I_{{D}^*_s}$=$I_{\bar{D}^*_s}$=1. The peak near
3960~MeV is analogous to Fig.~\ref{Fig: 1} and Fig.~\ref{Fig: 2}, which also
favors the assumption of the $X(3960)$ as a $D_s\bar{D}_s$ molecular state,
whereas a small bump is obtained instead of a dip at around 4140~MeV. This
bump-like structure in the $D_s^+{D}_s^-$ invariant mass spectrum is caused by
the $D_s\bar{D}_s$-$J/\psi\phi$ coupling. However, we cannot reproduce a dip
from this hump-like structure by adjusting the parameters. In the current model,
the coupled-channel effect with coupling to the $J/\psi\phi$ channel is
constrained by the heavy quark symmetry as shown in the Lagrangians.  In the
experimental treatment by the LHCb collaboration~\cite{LHCb:2022vsv}, the free
parameters were introduced to adjust the contributions of the coupled-channel
effect. Hence, our results do not conflict with experimental analysis, but are
obtained with more theoretical constraint.

Besides, adjusting other parameters,  such as the value of flavor factor, doesn't
work either. To prove this point, we take values of flavor
factor (FF) of the exchanged mesons $I_{D_s}$, $I_{\bar{D}_s}$, $I_{{D}^*_s}$ and
$I_{\bar{D}^*_s}$ in this process all equal to 0.1, 0.5, 1, 1.5 and 2,
respectively. From Fig.~\ref{Fig: 3} we can see although the value of the peak
for the bump-like structure around 4140~MeV gradually decrease with the
decreasing of the value of FF, a dip structure still cannot be reproduced in our
model. Based on these calculations, one can say that, given the fact that the coupled
channel $D_s\bar{D}_s$-$\phi J/\psi$  cannot reproduce a dip around
4140~MeV in our model, the $X_0(4140)$ could not be easily interpreted as a
$D_s\bar{D}_s$-$\phi J/\psi$ coupling. 

\section{Summary}\label{sec6}

Inspired by the newly observed $X(3960)$, we study the $B$ meson decay with
$D_s\bar{D}_s$ rescattering and couplings to channels $D\bar{D}$  and
$J/\psi\phi$  in a qBSE approach with spin parity $0^+$.  With the help of
effective Lagrangian, the potential kernel can be constructed by meson exchanges
and the invariant mass spectra are calculated and compared with the LHCb
experiment.

With the $D_s\bar{D}_s$ rescattering, a pole is found at 3932.8~MeV on the real
axis near $D_s\bar{D}_s$ threshold,  which shows that a very weakly  bound state
can be produced from the $D_s\bar{D}_s$ interaction. A peak structure near the
$D_s\bar{D}_s$ threshold is found in the $D_s^+{D}_s^-$  invariant mass spectrum
and have a good agreement with the experimental results, favoring the assumption
of the $X(3960)$ as a $D_s\bar{D}_s$ molecular state. After adding $D\bar{D}$ as
intermediate and final channel in the process, the theoretical $D_s^+{D}_s^-$ invariant mass
spectrum is still in agreement with the relevant experimental data. 

After adding $D\bar{D}$ channel, an extremely sharp peak can be obtained near
3930~MeV in $D^+{D}^-$ invariant mass spectrum. The $\chi_{c0}(3930)$ can be
assigned as the same $D_s\bar{D}_s$ state in the $D_s^+{D}_s^-$  invariant mass
spectrum, but with a very small width. The experimental structure around
3930~MeV in the $D^+{D}^-$ invariant mass spectrum can not be reproduced in the
current model only with this molecular state with $0^{++}$ from the
$D_s\bar{D}_s$ interaction.  The experimental analysis of such structure
suggests that it may be formed by states with spins 0 and 2, $\chi_{c0}(3930)$
and $\chi_{c2}(3930)$. The current results indict that the $\chi_{c2}(3930)$
may play a more important role  to form the structure near 3930~MeV in the
$D^+{D}^-$ invariant mass spectrum~\cite{LHCb:2020pxc}. 

With an additional Breit-Wigner resonance, the dip can be well reproduced around
4140~MeV in the $D_s^+{D}_s^-$ invariant mass spectrum. The possible role of the
$D_s\bar{D}_s$-$J/\psi\phi$  coupled-channel effect on the appearance of the dip
$X(4140)$ is also discussed in the current model.  However, a small bump instead
of a dip is found around 4140~MeV. Moreover, the dip can not be reproduced by
adjusting the parameters, which does not support the  assignment of  $X_0(4140)$
as the $D_s\bar{D}_s$-$J/\psi\phi$ coupled-channel effect. It suggests that the
$X(4140)$ has an independent origin different from the peak $X(3960)$ which
can be well reproduced from the $D_s\bar{D}_s$ interaction in the current work.

\vskip 10pt

\noindent {\bf Acknowledgement} Authors thank Dr. Bin Zhong for useful discussion. This project is supported by the National Natural Science
Foundation of China (Grants No. 11675228).

%

%\bibliography{../../../reference/Jabref/bibliography}

\begin{thebibliography}{23}%

%\cite{LHCb:2022vsv}
\bibitem{LHCb:2022vsv}
 [LHCb],
``Observation of a resonant structure near the $D_s^+ D_s^-$ threshold in the $B^+\to D_s^+ D_s^- K^+$ decay,''
[arXiv:2210.15153 [hep-ex]].
%18 citations counted in INSPIRE as of 28 May 2023

%\cite{Belle:2004lle}
\bibitem{Belle:2004lle}
K.~Abe \textit{et al.} [Belle],
``Observation of a near-threshold omega J/psi mass enhancement in exclusive $B \to K \omega J/\psi$ decays,''
Phys. Rev. Lett. \textbf{94}, 182002 (2005)
%doi:10.1103/PhysRevLett.94.182002
%[arXiv:hep-ex/0408126 [hep-ex]].
%524 citations counted in INSPIRE as of 28 May 2023

%\cite{Belle:2009and}
\bibitem{Belle:2009and}
S.~Uehara \textit{et al.} [Belle],
``Observation of a charmonium-like enhancement in the $\gamma \gamma  \to \omega J/\psi$ process,''
Phys. Rev. Lett. \textbf{104}, 092001 (2010)
%doi:10.1103/PhysRevLett.104.092001
%[arXiv:0912.4451 [hep-ex]].
%191 citations counted in INSPIRE as of 28 May 2023

%\cite{BaBar:2012nxg}
\bibitem{BaBar:2012nxg}
J.~P.~Lees \textit{et al.} [BaBar],
``Study of $X(3915) \to J/\psi \omega$ in two-photon collisions,''
Phys. Rev. D \textbf{86}, 072002 (2012)
%doi:10.1103/PhysRevD.86.072002
%[arXiv:1207.2651 [hep-ex]].
%118 citations counted in INSPIRE as of 28 May 2023

%\cite{LHCb:2020pxc}
\bibitem{LHCb:2020pxc}
R.~Aaij \textit{et al.} [LHCb],
``Amplitude analysis of the $B^+\to D^+D^-K^+$ decay,''
Phys. Rev. D \textbf{102}, 112003 (2020)
%doi:10.1103/PhysRevD.102.112003
%[arXiv:2009.00026 [hep-ex]].
%165 citations counted in INSPIRE as of 28 May 2023


%\cite{ParticleDataGroup:2022pth}
\bibitem{ParticleDataGroup:2022pth}
R.~L.~Workman \textit{et al.} [Particle Data Group],
``Review of Particle Physics,''
PTEP \textbf{2022}, 083C01 (2022)
%doi:10.1093/ptep/ptac097
%1093 citations counted in INSPIRE as of 31 May 2023

%\cite{Mutuk:2022ckn}
\bibitem{Mutuk:2022ckn}
H.~Mutuk,
``Molecular interpretation of X(3960) as $D_s^+ D_s^-$ state,''
Eur. Phys. J. C \textbf{82}, no.12, 1142 (2022)
%doi:10.1140/epjc/s10052-022-11120-3
%[arXiv:2211.14836 [hep-ph]].
%7 citations counted in INSPIRE as of 28 May 2023

%\cite{Xin:2022bzt}
\bibitem{Xin:2022bzt}
Q.~Xin, Z.~G.~Wang and X.~S.~Yang,
``Analysis of the X(3960) and related tetraquark molecular states via the QCD sum rules,''
AAPPS Bull. \textbf{32}, no.1, 37 (2022)
%doi:10.1007/s43673-022-00070-3
%[arXiv:2207.09910 [hep-ph]].
%17 citations counted in INSPIRE as of 28 May 2023

%\cite{Ji:2022uie}
\bibitem{Ji:2022uie}
T.~Ji, X.~K.~Dong, M.~Albaladejo, M.~L.~Du, F.~K.~Guo and J.~Nieves,
``Establishing the heavy quark spin and light flavor molecular multiplets of the X(3872), Zc(3900), and X(3960),''
Phys. Rev. D \textbf{106}, no.9, 094002 (2022)
%doi:10.1103/PhysRevD.106.094002
%[arXiv:2207.08563 [hep-ph]].
%15 citations counted in INSPIRE as of 28 May 2023

%\cite{Xie:2022lyw}
\bibitem{Xie:2022lyw}
J.~M.~Xie, M.~Z.~Liu and L.~S.~Geng,
``Production rates of Ds+Ds- and DD molecules in B decays,''
Phys. Rev. D \textbf{107}, no.1, 016003 (2023)
%doi:10.1103/PhysRevD.107.016003
%[arXiv:2207.12178 [hep-ph]].
%14 citations counted in INSPIRE as of 28 May 2023



%\cite{Bayar:2022dqa}
\bibitem{Bayar:2022dqa}
M.~Bayar, A.~Feijoo and E.~Oset,
``X(3960) seen in Ds+Ds- as the X(3930) state seen in D+D-,''
Phys. Rev. D \textbf{107}, no.3, 034007 (2023)
%doi:10.1103/PhysRevD.107.034007
%[arXiv:2207.08490 [hep-ph]].
%11 citations counted in INSPIRE as of 28 May 2023


%\cite{Chen:2023eix}
\bibitem{Chen:2023eix}
Y.~Chen, H.~Chen, C.~Meng, H.~R.~Qi and H.~Q.~Zheng,
``On the nature of X(3960),''
Eur. Phys. J. C \textbf{83}, no.5, 381 (2023)
%doi:10.1140/epjc/s10052-023-11527-6
%[arXiv:2302.06278 [hep-ph]].
%2 citations counted in INSPIRE as of 12 Jun 2023

%\cite{Badalian:2023qyi}
\bibitem{Badalian:2023qyi}
A.~M.~Badalian and Y.~A.~Simonov,
``The scalar exotic resonances $X(3915), X(3960), X_0(4140)$,''
Eur. Phys. J. C \textbf{83}, no.5, 410 (2023)
%doi:10.1140/epjc/s10052-023-11590-z
%[arXiv:2301.13597 [hep-ph]].
%3 citations counted in INSPIRE as of 28 May 2023

%\cite{Agaev:2023gti}
\bibitem{Agaev:2023gti}
S.~S.~Agaev, K.~Azizi and H.~Sundu,
``Near-threshold structures in the Ds+Ds- mass distribution of the decay B+Ds+Ds-K+,''
Phys. Rev. D \textbf{107}, no.9, 094018 (2023)
%doi:10.1103/PhysRevD.107.094018
%[arXiv:2303.02457 [hep-ph]].
%0 citations counted in INSPIRE as of 28 May 2023



%\cite{Ding:2021igr}
\bibitem{Ding:2021igr}
Z.~M.~Ding, H.~Y.~Jiang, D.~Song and J.~He,
``Hidden and doubly heavy molecular states from interactions $D^{(*)}_{(s)}{{\bar{D}}}^{(*)}_{s}$/$B^{(*)}_{(s)}{{\bar{B}}}^{(*)}_{s}$ and ${D}^{(*)}_{(s)}D_{s}^{(*)}$/${B}^{(*)}_{(s)}B_{s}^{(*)}$,''
Eur. Phys. J. C \textbf{81}, no.8, 732 (2021)
%doi:10.1140/epjc/s10052-021-09534-6
%[arXiv:2107.00855 [hep-ph]].
%22 citations counted in INSPIRE as of 28 May 2023

%\cite{Braaten:2004fk}
\bibitem{Braaten:2004fk}
E.~Braaten, M.~Kusunoki and S.~Nussinov,
``Production of the X(3870) in B meson decay by the coalescence of charm mesons,''
Phys. Rev. Lett. \textbf{93}, 162001 (2004)
%doi:10.1103/PhysRevLett.93.162001
%[arXiv:hep-ph/0404161 [hep-ph]].
%75 citations counted in INSPIRE as of 31 May 2023

%\cite{LHCb:2022dvn}
\bibitem{LHCb:2022dvn}
 [LHCb],
``First observation of the $B^+ \rightarrow D_s^+ D_s^- K^+$ decay,''
Phys. Rev. Lett. \textbf{108} (2023), 034012
%doi:10.1103/PhysRevD.108.034012
%[arXiv:2211.05034 [hep-ex]].
%5 citations counted in INSPIRE as of 21 Aug 2023

 %\cite{He:2014nya}
\bibitem{He:2014nya}
  J.~He,
  ``Study of the $B\bar{B}^*/D\bar{D}^*$ bound states in a Bethe-Salpeter approach,''
  Phys.\ Rev.\ D {\bf 90}, 076008 (2014)
%  [arXiv:1409.8506 [hep-ph]].
  %%CITATION = ARXIV:1409.8506;%%
  %5 citations counted in INSPIRE as of 07 Oct 2015
%\cite{Chen:2014afa}
  %2 citations counted in INSPIRE as of 10 sept. 2015

%\cite{Cheng:1992xi}
\bibitem{Cheng:1992xi}
H.~Y.~Cheng, C.~Y.~Cheung, G.~L.~Lin, Y.~C.~Lin, T.~M.~Yan and H.~L.~Yu,
``Chiral Lagrangians for radiative decays of heavy hadrons,''
Phys. Rev. D \textbf{47}, 1030-1042 (1993)
%doi:10.1103/PhysRevD.47.1030
%[arXiv:hep-ph/9209262 [hep-ph]].
%170 citations counted in INSPIRE as of 31 May 2023

%\cite{Yan:1992gz}
\bibitem{Yan:1992gz}
T.~M.~Yan, H.~Y.~Cheng, C.~Y.~Cheung, G.~L.~Lin, Y.~C.~Lin and H.~L.~Yu,
``Heavy quark symmetry and chiral dynamics,''
Phys. Rev. D \textbf{46}, 1148-1164 (1992)
[erratum: Phys. Rev. D \textbf{55}, 5851 (1997)]
%doi:10.1103/PhysRevD.46.1148
%749 citations counted in INSPIRE as of 31 May 2023

%\cite{Wise:1992hn}
\bibitem{Wise:1992hn}
M.~B.~Wise,
``Chiral perturbation theory for hadrons containing a heavy quark,''
Phys. Rev. D \textbf{45}, no.7, R2188 (1992)
%doi:10.1103/PhysRevD.45.R2188
%852 citations counted in INSPIRE as of 31 May 2023

%\cite{Burdman:1992gh}
\bibitem{Burdman:1992gh}
G.~Burdman and J.~F.~Donoghue,
``Union of chiral and heavy quark symmetries,''
Phys. Lett. B \textbf{280}, 287-291 (1992)
%doi:10.1016/0370-2693(92)90068-F
%674 citations counted in INSPIRE as of 31 May 2023

%\cite{Casalbuoni:1996pg}
\bibitem{Casalbuoni:1996pg}
R.~Casalbuoni, A.~Deandrea, N.~Di Bartolomeo, R.~Gatto, F.~Feruglio and G.~Nardulli,
``Phenomenology of heavy meson chiral Lagrangians,''
Phys. Rept. \textbf{281}, 145-238 (1997)
%doi:10.1016/S0370-1573(96)00027-0
%[arXiv:hep-ph/9605342 [hep-ph]].
%653 citations counted in INSPIRE as of 05 Jun 2023 



%\cite{Falk:1992cx}
\bibitem{Falk:1992cx}
A.~F.~Falk and M.~E.~Luke,
%``Strong decays of excited heavy mesons in chiral perturbation theory,''
Phys. Lett. B \textbf{292}, 119-127 (1992)
%doi:10.1016/0370-2693(92)90618-E
%[arXiv:hep-ph/9206241 [hep-ph]].
%260 citations counted in INSPIRE as of 05 Jun 2023

%\cite{Isola:2003fh}
\bibitem{Isola:2003fh}
C.~Isola, M.~Ladisa, G.~Nardulli and P.~Santorelli,
``Charming penguins in B ---\ensuremath{>} K* pi, K(rho, omega, phi) decays,''
Phys. Rev. D \textbf{68}, 114001 (2003)
%doi:10.1103/PhysRevD.68.114001
%[arXiv:hep-ph/0307367 [hep-ph]].
%168 citations counted in INSPIRE as of 05 Jun 2023

%\cite{Liu:2009qhy}
\bibitem{Liu:2009qhy}
X.~Liu, Z.~G.~Luo, Y.~R.~Liu and S.~L.~Zhu,
``X(3872) and Other Possible Heavy Molecular States,''
Eur. Phys. J. C \textbf{61}, 411-428 (2009)
%doi:10.1140/epjc/s10052-009-1020-4
%[arXiv:0808.0073 [hep-ph]].
%214 citations counted in INSPIRE as of 05 Jun 2023

%\cite{Chen:2019asm}
\bibitem{Chen:2019asm}
R.~Chen, Z.~F.~Sun, X.~Liu and S.~L.~Zhu,
``Strong LHCb evidence supporting the existence of the hidden-charm molecular pentaquarks,''
Phys. Rev. D \textbf{100}, no.1, 011502 (2019)
%doi:10.1103/PhysRevD.100.011502
%[arXiv:1903.11013 [hep-ph]].
%168 citations counted in INSPIRE as of 31 May 2023

 %\cite{Aceti:2014uea}
\bibitem{Aceti:2014uea}
  F.~Aceti, M.~Bayar, E.~Oset, A.~Martinez Torres, K.~P.~Khemchandani, J.~M.~Dias, F.~S.~Navarra and M.~Nielsen,
  ``Prediction of an $I=1$ $D \bar D^*$ state and relationship to the claimed $Z_c(3900)$, $Z_c(3885)$,''
  Phys.\ Rev.\ D {\bf 90} (2014) no.1,  016003
 % doi:10.1103/PhysRevD.90.016003
 % [arXiv:1401.8216 [hep-ph]].
  %%CITATION = doi:10.1103/PhysRevD.90.016003;%%
  %42 citations counted in INSPIRE as of 09 Dec 2017 %34 citations counted in INSPIRE as of 09 Dec 2017 

%\cite{He:2015mja}
\bibitem{He:2015mja}
  J. He,
  ``The $Z(3900)$ as a resonance from the $D\bar{D}^*$ interaction,''
  Phys.\ Rev.\ D,  {\bf 92}: 034004 (2015)
%  [arXiv:1505.05379 [hep-ph]].
  %%CITATION = ARXIV:1505.05379;%%
  %2 citations counted in INSPIRE as of 31 Aug 201

%\cite{Oh:2000qr}
\bibitem{Oh:2000qr}
Y.~s.~Oh, T.~Song and S.~H.~Lee,
``J / psi absorption by pi and rho mesons in meson exchange model with anomalous parity interactions,''
Phys. Rev. C \textbf{63}, 034901 (2001)
%doi:10.1103/PhysRevC.63.034901
%[arXiv:nucl-th/0010064 [nucl-th]].
%212 citations counted in INSPIRE as of 31 May 2023

 %\cite{He:2019ify}
\bibitem{He:2019ify}
J.~He,
``Study of $P_c(4457)$, $P_c(4440)$, and $P_c(4312)$ in a quasipotential Bethe-Salpeter equation approach,''
Eur. Phys. J. C \textbf{79}, no.5, 393 (2019)
%doi:10.1140/epjc/s10052-019-6906-1
%[arXiv:1903.11872 [hep-ph]].
%133 citations counted in INSPIRE as of 31 May 2023

%\cite{Ding:2020dio}
\bibitem{Ding:2020dio}
Z.~M.~Ding, H.~Y.~Jiang and J.~He,
``Molecular states from $D^{(*)}\bar{D}^{(*)}/B^{(*)}\bar{B}^{(*)}$ and $D^{(*)}D^{(*)}/\bar{B}^{(*)}\bar{B}^{(*)}$ interactions,''
Eur. Phys. J. C \textbf{80} (2020) no.12, 1179
%doi:10.1140/epjc/s10052-020-08754-6
%[arXiv:2011.04980 [hep-ph]].
%28 citations counted in INSPIRE as of 20 Aug 2023

%\cite{He:2017lhy}
\bibitem{He:2017lhy}
J.~He and D.~Y.~Chen,
``$Z_c(3900)/Z_c(3885)$ as a virtual state from $\pi J/\psi-\bar{D}^*D$ interaction,''
Eur. Phys. J. C \textbf{78}, no.2, 94 (2018)
%doi:10.1140/epjc/s10052-018-5580-z
%[arXiv:1712.05653 [hep-ph]].
%20 citations counted in INSPIRE as of 31 May 2023


 

 


  %\cite{He:2015yva}
\bibitem{He:2015yva}
  J.~He,
  ``Internal structures of the nucleon resonances N(1875) and N(2120),''
  Phys.\ Rev.\ C {\bf 91}, 018201 (2015)
 % doi:10.1103/PhysRevC.91.018201
 % [arXiv:1501.00522 [nucl-th]].
  %%CITATION = doi:10.1103/PhysRevC.91.018201;%%
  %8 citations counted in INSPIRE as of 09 Apr 2017
 
%\cite{He:2017aps}
\bibitem{He:2017aps}
  J.~He,
  ``Nucleon resonances $N(1875)$ and $N(2100)$ as strange partners of LHCb pentaquarks,''
  Phys.\ Rev.\ D {\bf 95} (2017) no.7,  074031
%  doi:10.1103/PhysRevD.95.074031
%  [arXiv:1701.03738 [hep-ph]].
  %%CITATION = doi:10.1103/PhysRevD.95.074031;%%
  %5 citations counted in INSPIRE as of 09 Dec 2017


























\end{thebibliography}

\end{document}